\documentclass[twocolumn, 11pt]{article}

\usepackage[margin=1in]{geometry}
\usepackage{amsmath}
\usepackage{mathtools}
\usepackage{graphicx}
\usepackage[square,numbers]{natbib}
\usepackage{caption}
\usepackage{url}

\title{Convolutional Autoencoders for \\Lossy Light Field Compression}
\author{Svetozar Zarko Valtchev, Jianhong Wu}
\date{
Laboratory for Industrial and Applied Mathematics\\%
York University\\%
August 12, 2019
}

\begin{document}
\maketitle

\begin{abstract}

\textbf{
Expansion and reduction of a neural network's width has well known properties in terms of the entropy of the propagating information. When carefully stacked on top of one another, an encoder network and a decoder network produce an autoencoder, often used in compression. Using this architecture, we develop an efficient method of encoding and decoding 4D Light Field data, with a substantial compression factor at a minimal loss in quality. Our best results managed to achieve a compression of 48.6x, with a PSNR of 29.46 dB and a SSIM of 0.8104. Computations of the encoder and decoder can be run in real time, with average computation times of 1.62s and 1.81s respectively, and the entire network occupies a reasonable 584MB by today's storage standards. 
}
\end{abstract}

\section{Introduction}

As the demand for virtual reality, and immersive experiences rises in the coming years, new challenges continue to arise in the way we store, transfer and consume our digital content. One intrinsic problem that has always been present, is the nature in which file sizes and computational loads seems to always be a step beyond state-of-the-art hardware's capabilities. Consequently, a standard for the next wave of media files must be reached.

In this paper we will discuss the basics of a light fields and outline relevant compression and decompression research for the medium thus far. We will then propose our own method utilizing modern deep learning tools and practices for the entire encoding/decoding framework. Our network will be carefully crafted and fine-tuned as we deem important for real world application and adoption. We will cover the entire data gathering, preprocessing, training and testing stages, before finally comparing our results to other leading alternatives in the field.

\subsection{Immersive Reconstruction}

Multiview displays along with Volumetric Geometry Models are currently the clashing viewpoints for creating functional immersive content. Volumetric methods are based on building polygons in a closed environment and applying textures and shading tools to them before rendering them to the user, often used in video games, animated movies and medical imaging \cite{DBLP:journals/corr/MilletariNA16}. The common critiques of Volumetric based techniques are their inability to correctly account for viewer-position-dependent effects such as opacity and occlusion \cite{Volumetric1}, as well as the large bandwidth requirements due to the density of volumetric pixels (known as voxels) \cite{Volumetric2,Volumetric3,Volumetric4}. Some estimates have been made as high as 135GB/s for smooth video playback for such a system \cite{Volumetric5}.

Multiview displays on the other hand rely on synthesizing different views needed by the user, based on properties of light and the plenoptic function.

\subsection{Light Fields and the Plenoptic Function}\label{lightfields}
\begin{figure}
    \centering
    \includegraphics[scale=.8]{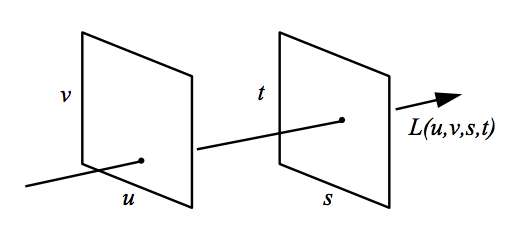}
    \caption{$uv$-$st$ plane parametrization of a Light Field as taken from Levoy et al. \cite{Levoy:1996:LFR:237170.237199}. The $uv$ plane represents a perspective location of an image of a scene analogous to the location of a photographer's camera, while the $st$ plane represents the spatial representation of the scene analogous to a still image taken the the camera. }
    \label{fig:lightSlab}
\end{figure}

The plenoptic function is a 5-D field representing the radiance of a ray of light, parameterized by the coordinates $(x,y,z)$, and the spherical angles $(\theta,\phi)$. However, the system happens to be overdetermined in this formulation, as radiance along a given ray of light remains constant. Using a carefully constructed parametrization, the system can be reduced down to 4 dimensions. This is known as a Light Field. There are many choices for the parametrization of a Light Field, (such as through 2 points on a unit sphere described entirely though $(\theta_1,\phi_1,\theta_2,\phi_2)$), but the most common is the pair of points on separate planes formulation. This is known as the $uv$, $st$ plane parametrization, where points on the $uv$ plane define the perspective (or viewpoint) of an image, and the $st$ the actual pixels in the image (or view). A visual explanation of this can be seen in Figure \ref{fig:lightSlab}.

Light Field Rendering became popularized at ACM SIGGRAPH 96' by Mark Levoy based on a talk and paper by the same name \cite{Levoy:1996:LFR:237170.237199}. Though computational means and displays have come a long way since, the overall idea is still relatively the same. Light Fields encompass all the information needed to synthetically reproduce a scene from all perspectives in a nearby neighbourhood of the observed ground truth. This can include viewpoints in-between those already measured on the $uv$ plane using techniques such a cubic interpolation, as well as viewpoints off the $uv$ plane, simulating viewer depth changes relative to the image.

\subsection{Compression and Autoencoders}

Compression generally splits into 2 categories: lossy and lossless. As the names imply, lossy compression allows for the lost of information, whereas lossless does not, at the cost of compression size. Lossless compression techniques are based on the idea that the data contains redundant information which can be packed more efficiently, while lossy relies on the removal of less important information present in the data. In this fashion, lossless compression allows for perfect reconstruction of the original data, while lossy has an error associated with it, commonly measured by the Mean Squared Error (MSE), Peak Signal-to-Noise Ratio (PSNR) \cite{qualityMeasures} or the Structural Similarity Index (SSIM) \cite{SSIM} for images.

With the rise of deep learning in recent years, a reasonable attempt at compression would see the utilization of an neural network. More precisely, the goal is to reduce the size of the initial data during the transfer/streaming and storage phases, while also limiting our data quality loss to a minimum, during the consumption phase. All this of course ideally, should be done fast enough to run in real-time. Hence, what's required is an efficient encoder and decoder. 

\begin{figure}
    \centering
    \includegraphics[scale=.33]{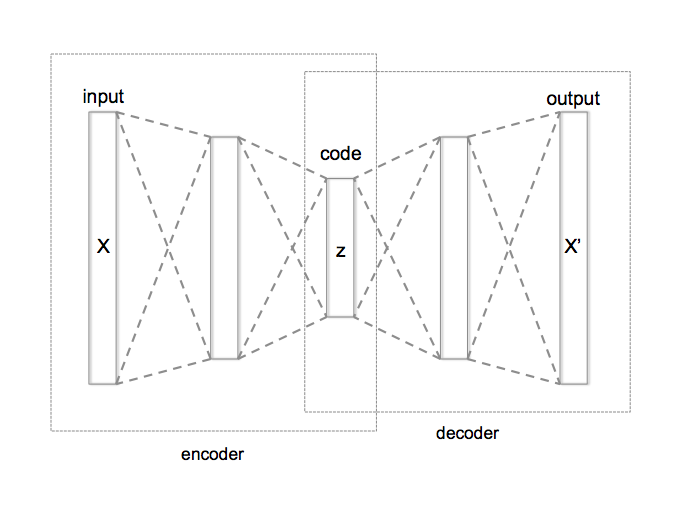}
    \caption{General structure of an autoencoder network \cite{wiki:1}}
    \label{fig:autoencoder}
\end{figure}

Generally, neural network designs that fan inwards layer to layer, tend to reduce down the amount of information required to represent the data. Alternatively, designs that fan out layer to layer tend to increase the information. The earlier structure is known as an encoder network, whereas the latter as a decoder network. Stacking these two one after the other, and training them together as one, gives rise to what is known as an autoencoder \cite{Hinton504}. A visual representation of this architecture can be seen in \ref{fig:autoencoder}.

An autoencoder network is a common tool for compression of data. They have achieved great results in the encoding of images \cite{imageCompression1, imageCompression2, imageCompression3}, videos \cite{videoCompression1} and synthesization \cite{audioCompression1} and style transfer of audio \cite{audioCompression2}.

The input and output of the network are the same data sample and ideally the network will figure out the optimal encoding in the hidden layers, by varying the weights of the model. The entire network gets trained together, but the encoder and decoder can be separated post-training, as to allow for separate encoding and decoding functions. The middle layer (output to encoder sub-network and input to decoder sub-network) contains a latent representation of the input/output data in the compression. This layer does not necessarily have an observable meaning associated with each neuron, but this is not a problem as applying the decoder network to it can map it back to a meaningful result.

The size of this middle layer relative to the input layer defines the compression size. This can be fine tuned as desired, based on the architecture design of the network. Of course there is a trade-off between compression size and compression loss, strongly dependant on this bottleneck layer. Too large of a layer keeps the compression low, but too small of a choice can lead to artifacts such as ringing, blocking, colour distortion, image blur \cite{compressionArtifacts}, and checkerboard patterns \cite{odena2016deconvolution}.

\subsection{Convolutional Neural Networks}

When it comes to network structures for visual data such as images and video, it has been shown that convolutional networks \cite{Fukushima1980} are best for extracting local spatial information in frames such as edge and corner detection \cite{DBLP:journals/corr/ZeilerF13, DBLP:journals/corr/YosinskiCNFL15}. The breakthrough itself is one of the major reasons deeplearning has rose to claim recently, through the success of networks such as LeNet, GoogleNet, VGGNet, ResNet and most notably AlexNet \cite{lenet, googlenet, vggnet, resnet, alexnet}. The convolution architecture also allows for naturally arising reduction in the size of layer to layer, with or without the use of pooling layers. This is useful in building the encoder subnetwork. For the decoder part however, an inverse type of convolution operation is required, known as deconvolution \cite{deconv}. Deconvolutional layers are not as common as convolutional ones, but their use is well understood as an "unpooling" agent in the network, well suited for a decoder.

\subsection{Light Field Compression}

With the relative youth of the technology, there still is little known about effective compression of light fields. The JPEG group has proposed standards when it comes to light representation imagery files, in the form of what they call JPEG Pleno \cite{Schelkens:262255}, similar to their other JPEG codecs for imagery. Little is still agreed upon, and much is still in debate in the community. 

From a lossless perspective, Santos et al. \cite{Santos2017LosslessLC} look at alternative data arrangements and colour transformations to achieve a more compact representation of the data, with respectable results. Like any lossless approach however, they are still orders of magnitude off from large scale adoption.

When perfect data retention can be sacrificed for significant size reduction, a few reasonable attempts have shown promise. Chen et al. \cite{8030107} were able to reduce bit-rates close to 50\% while maintaining reasonably high PSNR values, by carefully encoding disparity information with some optimized key views of the light field. Zhang et al. \cite{8546779} proposed an alternative method encoding a simulated point cloud into a B-Spline wavelet, achieving PSNR values near 30 dB depending on compression parameters, but their computation times vary between 5 and 24 seconds.

More similar to our inspiration, Barik et al. \cite{8451597} utilize the use of convolutional neural networks to achieve bitrate gains of 30\%. Barik proposes a standard video encoder to transform the light fields into sparse representation of the data, and a neural network to decode it back to the original light field. Computation times were not outlined. Gupta et al. \cite{8014902} go one step further and rely only on a neural network, splitting the spatial and angular encoding tasks in separate branches of their network, until combining the results in the final layers. Their model achieved respectable PSNR scores between 26-32 dB, attaining a compression ratio of 16:1, but coming at the cost of computation time, logging processing times on the order of minutes.

\section{Data}\label{data}
\subsection{Resources}

There are far less open source light field datasets available, than for more mainstream media content such as images and videos. The Stanford Light Field Archive is the most well known, with the datasets commonly showing up in research papers \cite{stanfordArchive}. One common issue with all the Stanford datasets for our purpose was the different sizes of supplied images. Images are often stretched in convolutional networks so as to have 1:1 ratios along their spatial dimensions, but this would provide problems in light filed images as the disparity maps would change along these stretched dimensions. The Old Archives resolutions are too low, where as the New Archives only provide 13 reasonable light fields, though they are made up of 289 viewpoints each. The Multiview and Three-View datasets each offered too large of baselines between viewpoints. 

MIT have their own achieve \cite{MITArchive}, with applications to other commonly cited papers \cite{Marwah:2013:CompressiveLightFieldPhotography}, but the data was entirely synthetic, lacking variations representative of true world structures such as materials, diffusion and lighting.

Lytro was a pioneer in the light field imaging sector, and strategically offer some publicly available data from captured on their own cameras \cite{Lytro}. They offer 25 high quality light fields, all in 1:1 image ratios, and great variety of viewpoints at close baselines, but the formatting and preprocessing needed to access the data, was far too convoluted. Furthermore, with the company closing down, and the format not being standardized we chose to go in a different direction.

We decide on the HCI Light Field Dataset as collected by groups from Heidelberg University and the University of Konstanz  \cite{LFData2016benchmark}. The dataset consists of 24 light fields, each made up of 81 viewpoints. All the image slices are of equal size and aspect ratios. This is ideal for an input layer of a neural network, as the size would have to be fixed, while the aspect ratio assures our images would not suffer from any resizing artifacts. Furthermore, the data already comes segmented into a test/train split ideal for research purposes, and the true disparity maps are also provided if we choose to utilize them in the future. Each light field is made up of 9x9 densely packed viewpoints, each of 512x512 pixels along 3 color channels, in standard .png format.

\subsection{Preprocessing}

Training neural networks requires vast amounts of data, yet we are limited by just 24 light fields. In order to provide our network a large enough sample size to train on we utilize a few data augmentation techniques commonly used for image datasets.

We first split our 24 light fields into testing and training partitions. Honauer et al. \cite{LFData2016benchmark} already proposed a testing standard of the first 4 light fields, so we use the subsequent 20 as our training set. We can double our sample size to 40 by allowing for horizontal flips of the light fields. Note however, that we abstain from flipping them vertically, as we would prefer our network to learn features in their correct rotation as oppose to generalized for different angles. 

\begin{figure}
    \centering
    \includegraphics[scale=.38]{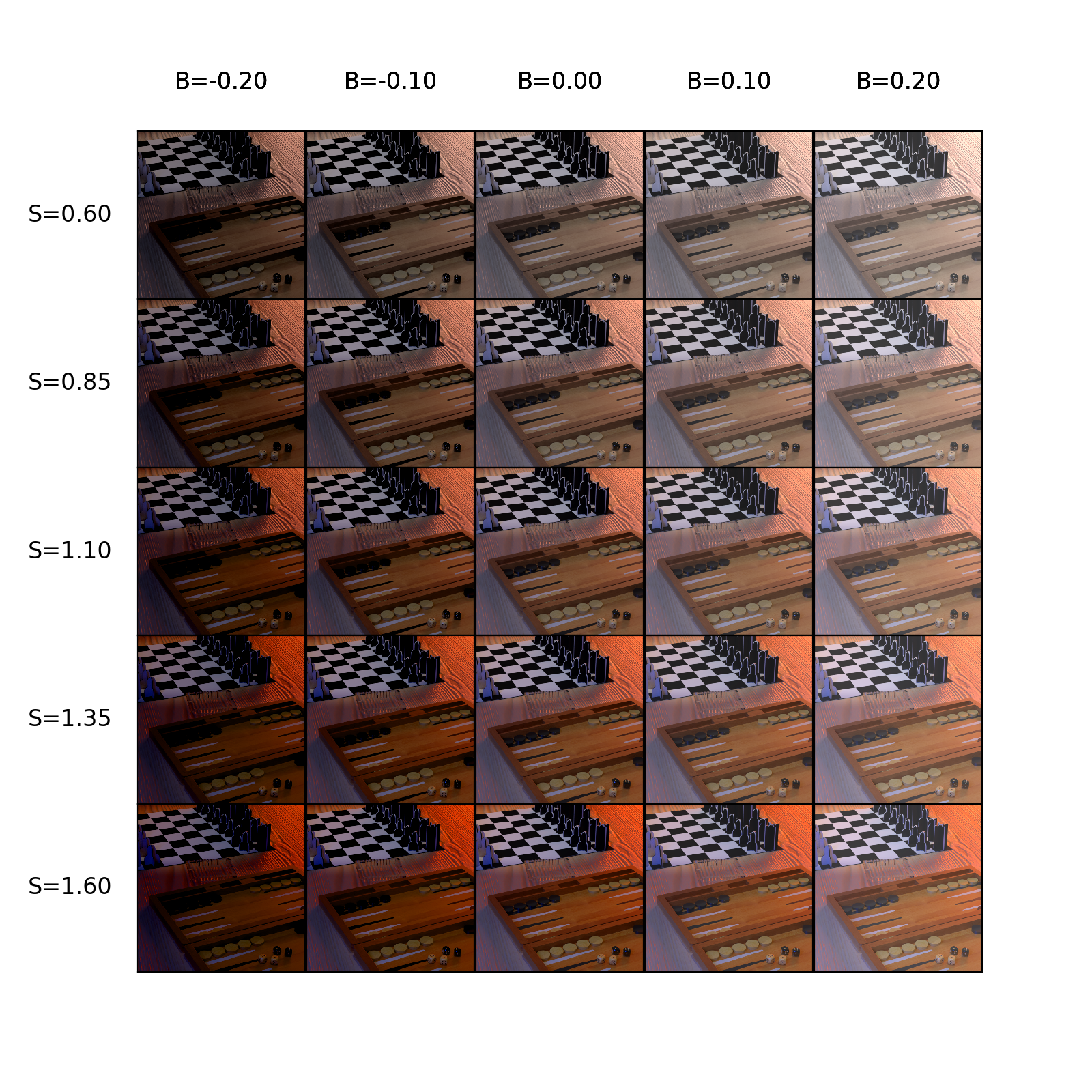}
    \caption{Adjusting brightness and saturation levels of a light field view to synthesize new data samples.}
    \label{fig:brightnessSaturation}
\end{figure}

\begin{figure}
    \centering
    \includegraphics[scale=.2]{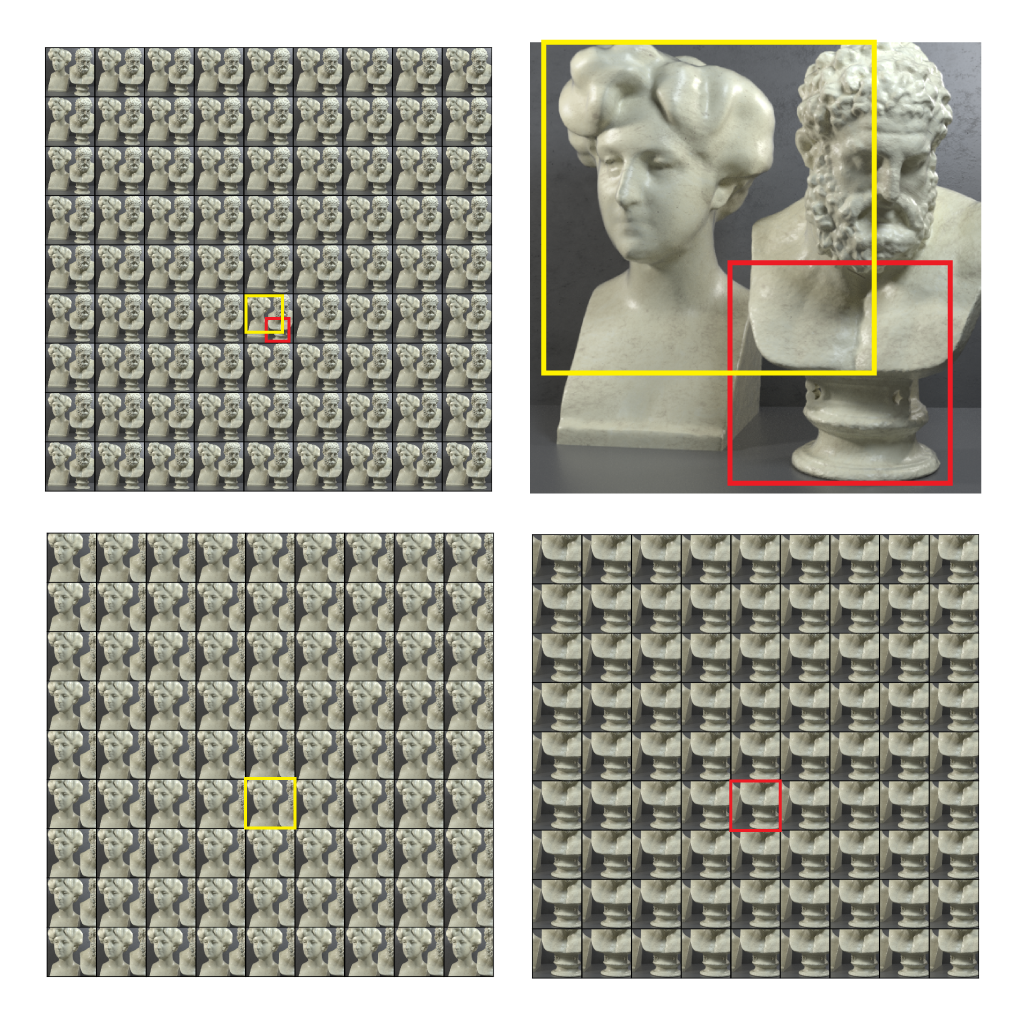}
    \caption{Beginning with our initial light field array, we can generate new light fields based on cropping and resizing different parts of the images. Top left is standard \textit{greek} light field, while bottom row is example of 2 new light fields created from the original.}
    \label{fig:subsampledLFs}
\end{figure}

Next, we proceed by applying random brightness and saturation changes to all viewpoints in a given light field to simulate 1000s of "new" light fields. We chose uniform random brightness adjustments $\in (-0.2, 0.2)$, while our saturation factors stay $\in (0.6, 1.6)$ (Note all our images are processed in float32 dtype, with pixel values ranging from 0 to 1). Lastly, we subsample viewpoints at random locations at random sized windows, keeping aspect ratios at 1:1 and resizing subimages back to the initial 512x512 size. The minimum size window we use for such subsampling is 256x256. This allows us to turn different parts of a light field into new light fields, at new zoom depths with new scales to their disparities maps, thus allowing for better generalization. Figures \ref{fig:subsampledLFs} and \ref{fig:brightnessSaturation} illustrate both of these techniques in more detail.

After all our augmentations, we have enough data to never see the exact same light field twice in a training period. Thus, the exact definition of a training epoch will be adjusted in section \ref{training}.

\section{Network Architecture}

To accomplish a significant compression rate, the architectural design of our network will be of utmost importance. More specifically the size of middle bottleneck layer relative to the input layer, will define our compression size by

$$
Compression\:Ratio = \frac{units\:in\:input\:layer}{units\:in\:middle\:layer}
$$

\noindent with, the middle layer being the layer that connects the encoder and decoder, and not necessarily the center most layer. The size of the input layer (and consequently the output layer) is entirely determined by the size of the full light field, as described in Section \ref{data}. The center-most viewpoint is most crucial for the network. This viewpoint will be used as a information highway \cite{highwayNetwork}, straight to the middle layer. The idea here is that there is high correlation between viewpoints of a light field, and so we can "hint" to our network to store a perfect copy of one viewpoint, and concentrate internally more on encoding the discrepancies between this image and the rest. The choice for using the center-most viewpoint is trivial as it is the one that is lies closest to all the rest from the Euclidean distance perspective, and therefore has the least variance between them. For example the parallax effects between the top-left and the bottom-right viewpoints will be far greater than 2 adjacent viewpoints, and therefore require more information to reconstruct one from the other.

The full light field will build up the other lane of the network, in a typical autoencoder-like structure. A typical light field can be structured as a 5 dimensional tensor, formally along the $u,v,s,t$ dimensions outlined in Section \ref{lightfields}, along with a 3 channel dimension standard in color imaging, for the red, green and blue light intensities in a given pixel. Instead of dealing with the tensor in this form decoupling information between the $uv$ and $st$ planes, we propose a stacking of all the viewpoints along the channel dimension. In this way we can apply precisely the common 2D Convolution techniques to the light field which is now reshaped to a high dimensional image. More importantly however, our convolutional filters will be applied to local regions across the $st$ spatial regions, and are inclined in theory to encoding disparity information directly.

We utilize convolutional layers with ReLU activations followed by batch normalization layers 5 times over, increasing the number of filters at each subsequent pair. All convolution layers utilize 2-dimension kernels of size 2, slid across the all spacial dimensions with stride lengths of 2 in all directions. At the end of the 5th normalization layer, we have the center viewpoint along with the encodings for reconstruction of the other views. This encoding can be thought of as having a latent part (disparity encodings) and an observable part (the center viewpoint). The network up to here is the encoder subnetwork. 

Our latent part is made up of a tensor of size 2048x16x16, while the observable part adds another 512x512x3 units, bringing the overall compression rate of the encoder to 48.6.

The decoder part, will look very similar to the encoder but in reverse. To "reverse" the convolutional layers of the encoder, we apply what is known as convolutional transpose layers (or deconvolutional layers) directly to the latent part of the encoded data. Once again we apply ReLU activations and batch normalization layers throughout, reversing the filter sizes of the convolutional layers we had before, to achieve a somewhat symmetric network.

\begin{figure*}
    \centering
    \includegraphics[scale=.4]{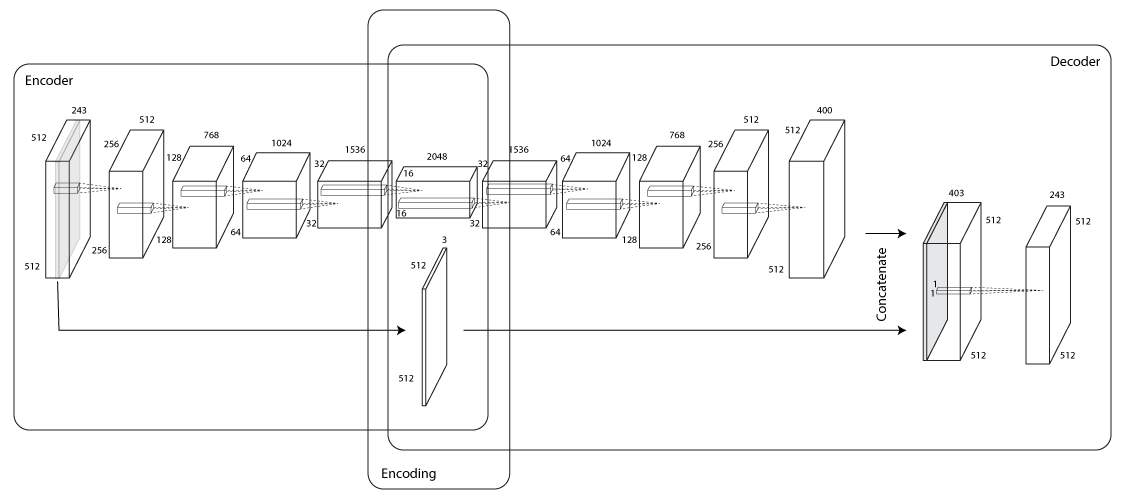}
    \caption{Full architecture of our Light Field Autoencoder. The network is comprised of 2 lanes. The top lane is a standard conv/deconv autoencoder, with all filter kernels having a size of 2x2, while the bottom is a network highway to pass information directly to the encoding and decoding stages. Finally the 2 lanes converge and the resultant tensor is trivially convoluted across each pixel separately, down the tensor depth, in order to reconstruct our initial input at the output layer.}
    \label{fig:LF_autoencoder}
\end{figure*}

Finally, we stack the observable part of our encoding to our resultant deconvolution, and apply one last trivial ReLU deconvolution layer with stride and kernel size of 1 (technically indistinguishable from a convolutional layer with these parameters), to get back our light field at the output layer. Each filter at this layer will give us back one channel of the initial light field stack, so obviously we require 243 filters, to make our inputs and our outputs equal. Therefore, one can abstract this layer as the final combination of the middle viewpoint plus the deviation from this image to all the others, through the propagated encoding of the network data up to this point. The detailed network architecture can be seen in Figure \ref{fig:LF_autoencoder}.

\section{Training}\label{training}

\begin{figure}
    \centering
    \includegraphics[scale=.5]{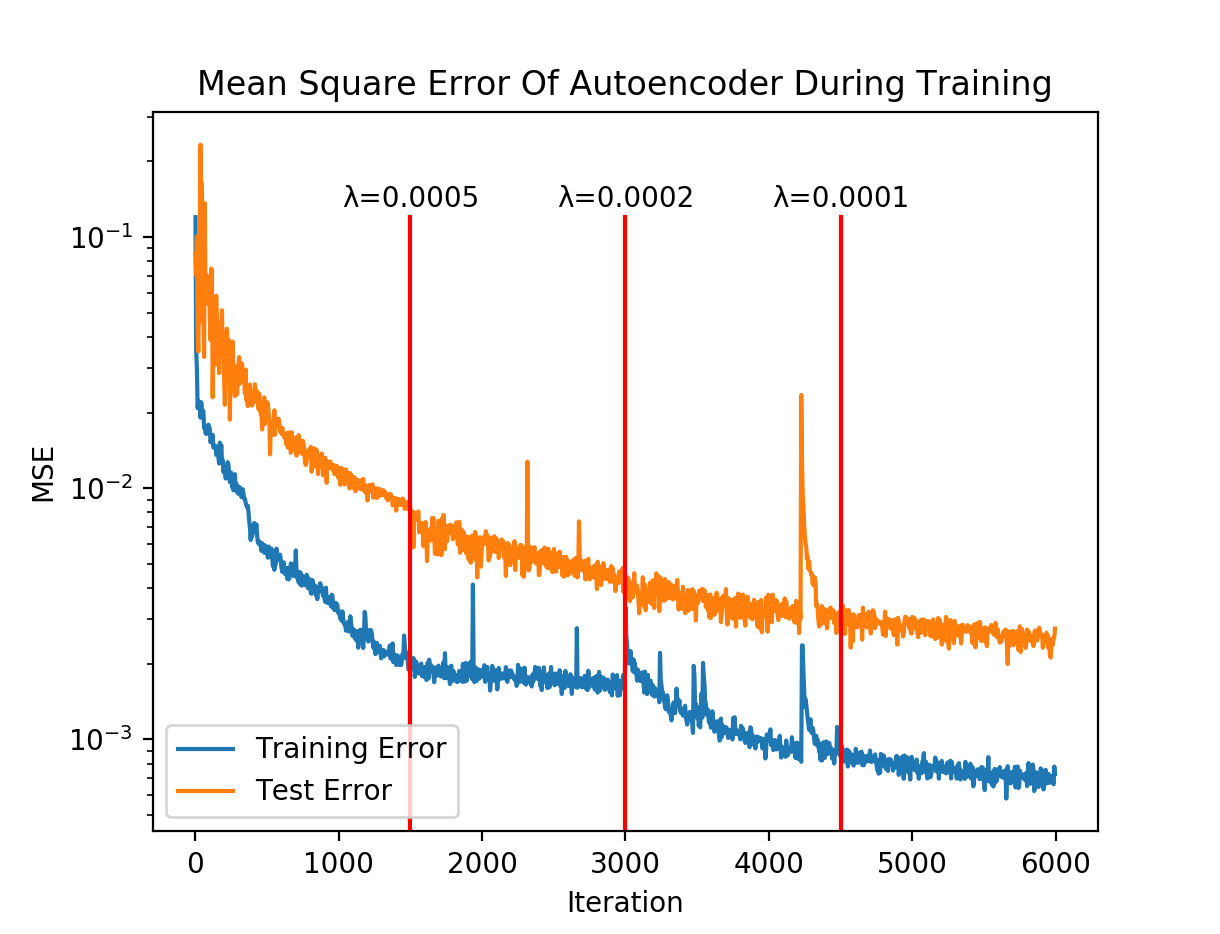}
    \caption{Mean squared error of the training and test set, vs training epoch. The red lines indicate learning rate parameter changes during the training phase.}
    \label{fig:error}
\end{figure}

Standard neural network training practices were followed for training our autoencoder. We utilize batch sizes of 4 light fields per iteration, as this was the limit of how much we could store in memory given the size of the network and the number of parameters associated with it. Each epoch consists of 30 iterations, finetuning the learning rate hyperparameter periodically based on error stabilization. More precisely, we found that lowering the learning rate every 30 epochs was reasonable, with our learning rate starting at 0.001, going down to 0.0005, 0.0002 and finally 0.0001. There were no huge error drops associated with the changes in the hyperparameter. See Figure \ref{fig:error} for more details. All training was done on a NVIDIA Tesla K80 on a standard cloud instance.

It is important to note that since we will not likely ever see the exact same light field twice, we never fully go through out entire dataset. Therefore, an epoch in our training will actually mean going through each of our initial light fields 300 times over, each generating a new synthetic light field for data variety purposes.

We utilize the commonly used ADAM optimizer in our weight updates during training. We set the loss function of our optimizer to reduce the mean square error between our input light field and the decoded light field at the end of the autoencoder (i.e difference between input and output layer to be minimized).

\section{Results}

Our autoencoder was left to train for 90 hours total, accounting for the numerous hyperparameter changes outlined in Section \ref{training}. Based on overall performance improvements stagnation, we decided to terminate the training at 6000 iterations. As can be seen in Figure \ref{fig:error}, we achieved an error rate of about 0.00062 for our training error, while our test error was 0.00229, an order of magnitude larger. Keep in mind, our error measurement at training was a simple Euclidean distance measure (MSE) between images. The exact error rates are presented in Table \ref{table:results}.

\begin{table}[]
\centering
\caption{Error Measurements of Test Data}
\begin{tabular}{|l|l|l|l|}
\hline
Sample  & MSE        & PSNR       & SSIM      \\ \hline
Origami & 0.0023536 & 26.28276 & 0.8104427 \\ \hline
Bicycle & 0.0034758 & 24.58939 & 0.6798532 \\ \hline
Herbs   & 0.0022105 & 26.55508 & 0.6492076 \\ \hline
Bedroom & 0.0011327 & 29.45878 & 0.7579887 \\ \hline
Mean    & 0.0022932 & 26.72150 & 0.7243731 \\ \hline
\end{tabular}
\label{table:results}
\end{table}

\begin{figure}
    \centering
    \includegraphics[scale=.18]{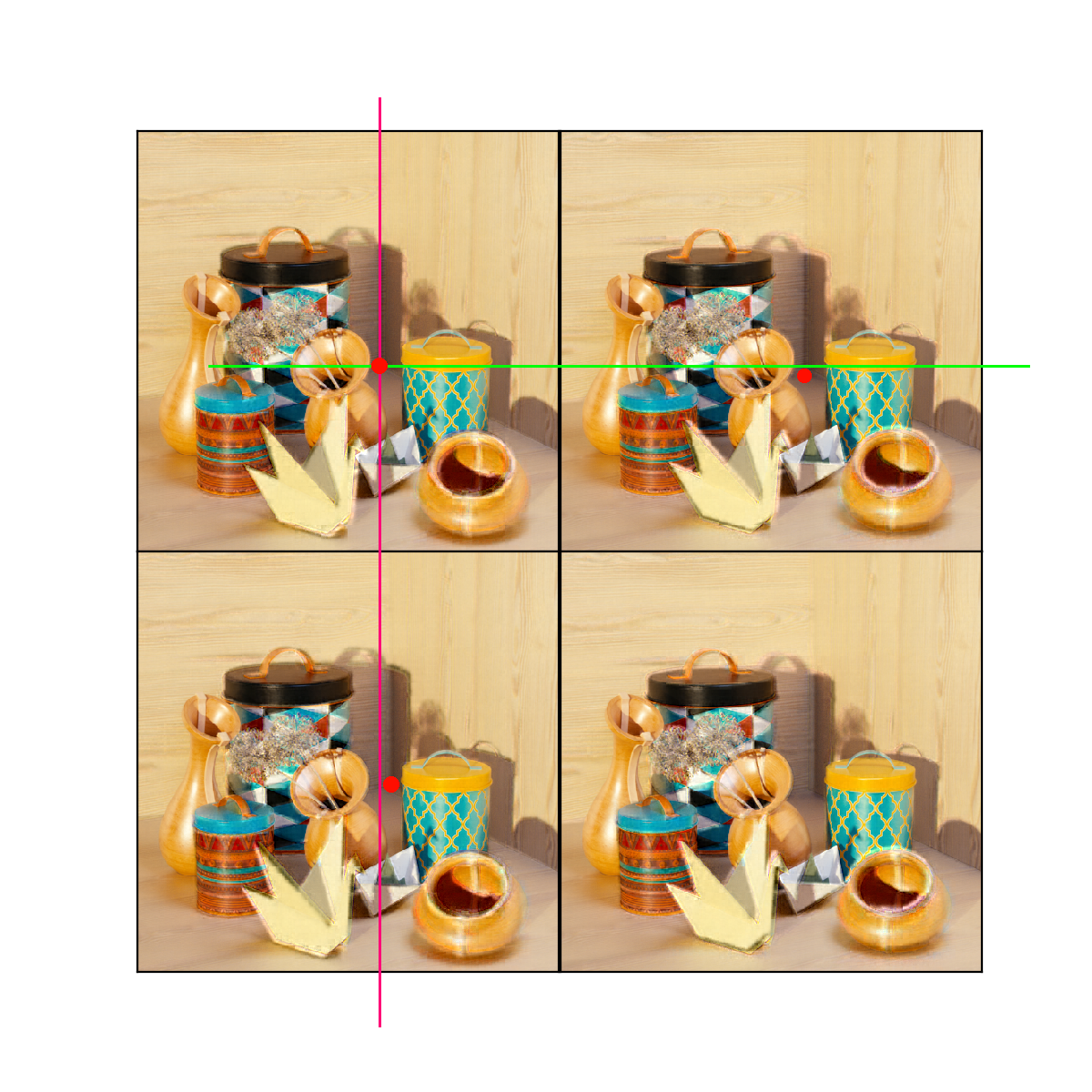}
    \caption{4 corner views of decoded Origami light field. Red dot in each image represents the corner of the walls in each view. The shift in adjacent views of this marker point is encouraging as it is visual proof of the encoding of the disparity of the light field.}
    \label{fig:results}
\end{figure}

\begin{figure}
    \centering
    \includegraphics[scale=.43]{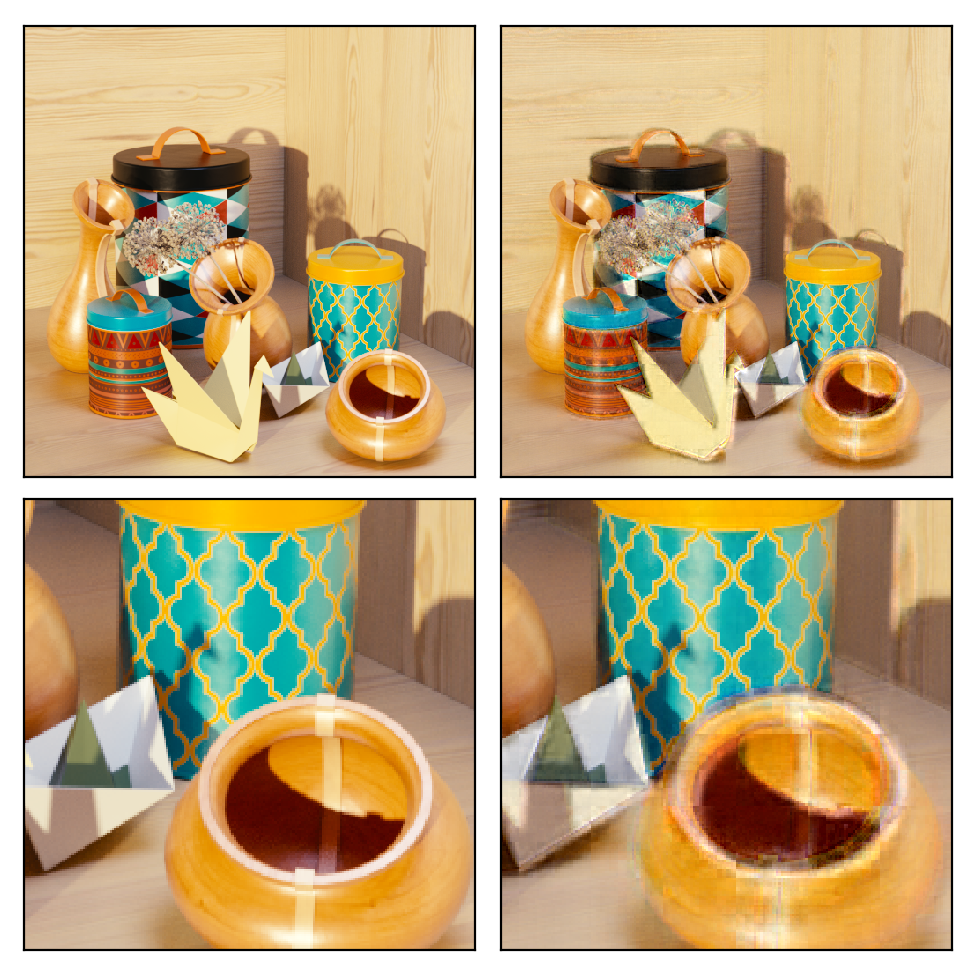}
    \caption{Visual comparison of original Origami light field and the decompressed light field using our network. (a) Images on the left show a single view of the light field, and a zoomed in section. Notice the sharpness and quality of the image at this level. (b) The same light field slab after compression and decompression. Note the blurring and blocking of the pixels in the zoomed in section around the edges of the pot and the shadows. }
    \label{fig:results2}
\end{figure}

Since our network designs relies on a very particular light field structure (512x512x9x9x3), a direct comparison with other methods is not trivial. Average PSNR values are in line with results reported from Gupta et al. \cite{8014902}, while achieving speed increases of 8x-130x depending on stride parameter choice. The compression rates reported by Gupta et al. are as low as 16:1, while we achieve a compression 3 times higher. Of course, our compression ratio is directly a function of the network architecture and cannot simply be fine tuned at run-time, so many of the measurements between our methods are not directly comparable. More precisely, given a trade-off between accuracy, speed and compression size, we opted towards a fixed high compression ratio of 48.6:1 and optimized our network for accuracy, while maintaining a moderate speed depending on the task at hand. Our processing time is again dependant on the depth, size and activation complexity of the network.

Despite the error rates being higher in the test set, there is cause for optimism. Actual disparity between different viewpoints is encoded quite well from a visual perspective. Figure \ref{fig:results} illustrates the 4 corners of a decoded light field using our autoencoder. Note the position of the corner of the wall, from one image to the next. This is precisely the parallax effect we would hope to preserve in the light field, and our autoencoder seems to do a good job of encoding this information. However, a more objective measure of this would be useful for further evaluation purposes. 

SSIM values seem to be in a reasonable range, though the SSIM also is not a perfect measure of visual quality of an image as perceived by humans. Collectively, the test light fields all clearly seem to produce visually recognizable views. However, it is clear that there is large compression artifacts present, largely in the form of blocking and blurring as can be seen in Figure \ref{fig:results2}. That being said, the artifacts seem to be localized to certain areas, textures and/or shapes, as can be seen in the edges of the pot compared to the pattern on the cylindrical box. These blockings are likely the cause for the high error values present in the MSE and PSNR measures.

The trained network occupies 588Mb of space, in line with state of the art neural networks like VGG16 and VGG19. Computation time for the end to end autoencoder was measured to be 3.617 seconds, with the encoder and decoder each taking approximately half the time. For the transfer of still light fields this is reasonable, however is impractical for high frequency transmission in it's current form.

\section{Discussion}

Despite some relatively high MSE values, we have implemented an autoencoder capable of encoding the 3-dimensional spatial information of a scene, with a compression ratio of 48.6:1. With our autoencoder, we can efficiently reduce a standard 25MB light field down to 0.5MB, and we can compute this compression in just under 2 seconds. Furthermore, we can decode this encoded information in another 2 seconds, making the sharing of light field scenes a reality. The entire network occupies just over half a GB, with the encoder and decoder each occupying approximately half the memory. Such a encoder/decoder system can easily and efficiently be implemented on the smallest of devices, without much problems.

The computation speed is a byproduct of the size of the overall network as well as the numerical data type as it propagates through the layers. Reducing the network size in terms of layers and/or units per layer should in theory greatly speed up the codec, but performance stability is not guaranteed. Float32 to Int8 conversion, along with weight quantization should greatly reduce the memory demand of the encoder, and could have more efficient calculation side effects. Weight and unit pruning could greatly reduce model size without a substantial loss to performance accuracy, and will likely be explored further. Alternatively, quantization and pruning can be applied only during inference, while the full network details are still kept for further training.

As with many lossy compression algorithms, our network did suffer from compression artifacts in the form of blurring, blocking and ringing. A possible explanation for the blurring, is likely due to our loss function. By optimizing for the MSE, like many other classifiers and regressors, our model tends to smooth out estimates. What instead we should aim for in visual content would be a sharper estimate as oppose to a more "median" fit to the training data. Further work into loss functions for images and light fields needs to be explored. Closely related is the concept of similarity between visual data. The SSIM we used above is one common standard, but still has discrepancies with actual qualitative human perception measures. Optimizing for SSIM is also an interesting idea that warrants further work.

\section*{Acknowledgements}
This work has been supported by the Natural Sciences and Engineering Research Council of Canada, and by the Canada Research Chairs program.

\bibliographystyle{unsrt}
\bibliography{master}

\newpage

\end{document}